\documentclass[notitlepage,superscriptaddress]{revtex4}

\usepackage{amscd,amsmath,amssymb}
\usepackage[usenames, dvipsnames]{color}

\newcommand{\p}[1]{(\ref{#1})}

\newcommand{\bQ}{{\overline Q}{}}

\newcommand{\bU}{{\overline U}{}}

\newcommand{\bpsi}{{\bar\psi}{}}

\newcommand{\be}{\begin{equation}}
\newcommand{\ee}{\end{equation}}
\newcommand{\bea}{\begin{eqnarray}}
\newcommand{\eea}{\end{eqnarray}}
\def\im{{\rm i}}
\renewcommand{\b}[1]{\bar{#1}}

\newcommand{\nn}{\nonumber}

\begin{document}
\title{Geometry and integrability  in $\mathcal{N}=8$ supersymmetric mechanics}

\author{Sergey Krivonos}
\email{krivonos@theor.jinr.ru}
\affiliation{Bogoliubov  Laboratory of Theoretical Physics, JINR,
141980 Dubna, Russia}
\author{Armen Nersessian}
\email{arnerses@yerphi.am}
\affiliation{Yerevan Physics Institute, Alikhanian Br. St. 2, 0036 Yerevan, Armenia}
\affiliation{Yerevan State University, 1 Alex Manoogian St., Yerevan, 0025, Armenia}
\affiliation{Bogoliubov  Laboratory of Theoretical Physics, JINR,
141980 Dubna, Russia}
\author{Hovhannes Shmavonyan}
\email{hovhannes.shmavonyan@yerphi.am}
\affiliation{Yerevan Physics Institute, Alikhanian Br. St. 2, 0036 Yerevan, Armenia}

\begin{abstract}\noindent
We construct the $\mathcal{N}=8$ supersymmetric mechanics  with potential term  whose configuration space is the  special K\"ahler manifold of rigid type and show that it can be viewed as the K\"ahler  counterpart  of $\mathcal{N}=4$ mechanics  related to ``curved WDVV equations''.
Then, we consider the special case of the  supersymmetric mechanics with the non-zero potential term defined on the family  of $U(1)$-invariant one-(complex)dimensional special  K\"ahler metrics. The bosonic parts of these  systems include
superintegrable deformations of  perturbed two-dimensional  oscillator and Coulomb systems.
\end{abstract}
\maketitle

\section{Introduction}
The construction of $\mathcal{N}$-extended supersymmetric mechanics  has remained one of the main directions 
of supersymmetric community since the introduction of the concept of supersymmetry. Nevertheless, until now there has been  no regular way to find the $\mathcal{N}>2$ supersymmetric extensions of the given mechanical systems.
The traditional  way to increase the number of supersymmetries (without exceeding the number of fermionic degrees of freedom) is to provide the configuration space with  the complex structure(s)(with appropriate specification of the potential term), i.e. to restrict the configuration space to  K\"ahler, hyper-K\"ahler or quaternionic manifold.
Say, on the  generic $N$-dimensional  Riemann manifold
one can always construct the  $\mathcal{N}=2$ supersymmetric mechanics  with  $(N|2N)$ (i.e. $N$ bosonic and $2N$ fermionic) degrees of freedom; requiring that configuration space be  generic $N$-(complex)dimensional K\"ahler manifold and properly specifying the potential,
we can    construct   the $\mathcal{N}=4$ supersymmetric mechanics with $(N|2N)$ (complex) degrees of freedom. The bosonic part of these Hamiltonians reads
\be
H_{(2)}=\frac12 g^{ij}(x)\left(p_ip_j+\partial_i W(x) \partial_j W(x)\right) \qquad \longrightarrow \qquad H_{(4)}=\frac12 g^{\bar a b}(z,\bar z)\left(\bar\pi_a\pi_b+\bar\partial_a \bar U(\bar z) \partial_b U(z)\right),
\label{1}\ee
with $(p_i, x^i)$ , $(\pi_a, z^a)$ being canonically conjugated pairs and $g^{ij}(x)$ and  $g^{\bar a b}$ being  the inverse Riemann and K\"ahler metrics, respectively.

Another way to increase  the number of supersymmetries (above $\mathcal{N}=2$ supersymmetry) is  the doubling  of fermionic degrees of freedom, supplied with introducing of additional geometric objects.
Say, to construct the  $\mathcal{N}=4$ supersymmetric extension of free-particle system on generic configuration space we have to
double the number of fermionic degrees of freedom from $2N$ to $4N$ and
 introduce the third-rank symmetric tensor $F_{ijk}(x)dx^idx^jdx^k$ which satisfies the  {\sl curved WDVV equations}  \cite{WDVVShort}
\be
  F_{kmj;i}= F_{kmi;j}, \quad
F_{jkp}g^{pq}F_{imq}-F_{ikp}g^{pq}F_{jmq}=R_{ijkm},
\label{mWDVV0}\ee
where $R_{ijkl}$  are  the components of  Riemann tensor of  $(M_0, g_{ij}dx^idx^j)$, and the subscript $;$ denotes
 covariant derivative with Levi-Civita connection.

Similarly, to construct the  $\mathcal{N}=8$ supersymmetric extension of free-particle system on K\"ahler manifold  we have to
increase the number of the (real) fermionic variables  from $4N$ to $8N$ and
 introduce  the third-rank holomorphic symmetric tensor  $f_{abc}(z)dz^adz^bdz^c$ which satisfies  the equations \cite{BKN}
\be
  f_{abc;d}= f_{abd;c}\;,
\qquad R_{a\bar b c\bar d}=-
f_{ace}g^{{\bar e}' e}{\bar f}_{\bar e'\bar b\bar d},
\label{sKrt}\ee
 where $f_{abc;d}= f_{abc,d}-\Gamma_{ad}^e f_{ebc}-\Gamma_{bd}^e f_{eac}-\Gamma_{cd}^e f_{eab}$,
and $R_{a\b{b}c\b{d}}$,   $\Gamma^a_{bc}$ 
are   the non-zero components of the Riemann tensor  and Levi-Civita connection which are defined by the relations
\be\label{11}
 \Gamma^a_{bc} =g^{a\b{d}}g_{b\b{d},c}, \qquad  R_{a\b{b}c\b{d}}=g_{n\b{b}} \left( \Gamma^n_{ac}\right){}_{, \b{d}}.
\ee
These manifolds are known as  the special K\"ahler manifolds of the rigid type \cite{fre} and they have been  extensively studied since their
introduction within the context of Seiberg-Witten duality \cite{SW}.
The  similarity between these systems  has  been not noticed before.

In this paper  we show that    this similarity holds  for the supersymmetric mechanics with the potential term as well.
Namely, after reviewing the main properties of $\mathcal{N}=4$ supersymmetric mechanics  connected with the solution of modified WDVV equations \cite{WDVVShort,2,kozyrev}({\sl Section 2}),  we construct on the special  K\"ahler manifold of the rigid type, the $\mathcal{N}=8$ supersymmetric mechanics with potential term ({\sl Section 3}).
We find that  for the doubling of the supersymmetries the prepotentials $W(x), U(z) $
in the bosonic Hamiltonians \eqref{1} should satisfy the following equations
\be
W_{i;j} + F_{ijk} g^{km} W_{m} =0,\qquad U_{a;b}-f_{abc}g^{\b{d}c}\bU_{\b{d}}=0.
\label{WF}\ee
Finally, we  present in the {\sl Section 3}  the general solution of the one-(complex)dimensional $U(1)$-symmetric special K\"ahler manifold  and find the admissible set of potentials
for $\mathcal{N}=8$ supersymmetric mechanics.
{ The bosonic parts of these supersymmetric mechanics include the superintegrable  perturbations of deformed two-dimensional oscillator and Coulomb system suggested in \cite{spainosc,spaincoul}}.

\setcounter{equation}{0}
\section{$\mathcal{N}=4$ mechanics on  Riemann manifolds}
In order to construct  the $\mathcal{N}=4$ supersymmetric mechanics
on $N$-dimensional   Riemannian manifold $(M_0, g_{ij}(x)dx^idx^j)$ we extend the cotangent bundle $(T^*M_0,dp_i\wedge dx^i$
by $4N$ fermionic variables  $\psi^{i\alpha}, \bpsi^j_{\beta }=\left( \psi^{\beta}_j\right)^\dagger$, with $su(2)$ indices $\alpha,\beta=1,2$
which are  raised and lowered as follows:
 $A_\alpha = \epsilon_{\alpha\beta}A^\beta,  A^\alpha=\epsilon^{\alpha\beta} A_\beta$ with $\epsilon_{12}=\epsilon^{21}=1$, and then
  define  the following transition maps from one chart to the other
\be
{\tilde x}^i={\tilde x}^i(x),\qquad {\tilde p}_i=\frac{\partial x^j}{\partial{\tilde x}^i}p_j,\qquad{\tilde\psi^{ia}}=\frac{\partial{\tilde x}^i(x)}{\partial x^j}\psi^{ja}.
\label{LT}\ee
Then we  equip this   supermanifold with  the supersymplectic structure  which is  manifestly  invariant with respect to  \eqref{LT}
\be\label{ss0}
\Omega=dp_i\wedge dx^i+
{\im}d\left(\psi^{i\alpha}g_{ij}{ D}\bpsi^{j}_{\alpha}- \bpsi^{i\alpha}g_{ij}{ D} \psi^{j}_{\alpha} \right)=  dp_i \wedge dx^i +
 \im R_{ijkl}\psi^{i\alpha}\bpsi^{j}_{\alpha} dx^k\wedge
dx^l +2\im g_{ij}D\psi^{i\alpha} \wedge D\bpsi^{j}_{\alpha},
\ee
where $D\psi^{i\alpha}\equiv d\psi^{i\alpha}+
\Gamma^i_{jk}\psi^{j\alpha} dx^k$, $\alpha=1,2$
and    $\Gamma^i_{jk}$, $R_{ijkl}$
 are the components of the connection and curvature
 of the metric
  $g_{ij}(x)dx^i dx^j$.

The Poisson brackets corresponding to this (super)symplectic structure,  are defined by the following nonzero relations
\be\label{PB}
\big\{{p}{}_j,x^i \big\}= \delta^{i}_j, \;\;\big\{ {p}_i, {\psi}^{j\alpha} \big\} = -\Gamma^j_{ik}{\psi}^{k\alpha},
 \;\; \big\{ {p}_i, { p}_j \big\} = 2\im R_{ijkm}{\psi}^{ k\alpha}{\bpsi}_{\alpha}^m ,\;\; \big\{ {\psi}^{i\alpha}, {\bpsi}_{\beta}^j    \big\} = -\frac{\im}{2}\delta^{\alpha}_{\beta} \, g^{ij}.
\ee

Our goal is to construct   the
supercharges $Q^\alpha, \bQ{}_\beta$,  and the Hamiltonian $\mathcal{H}$ which obey the  $\mathcal{N}=4, d=1$ Poincar\'{e} superalgebra
\be\label{N4Poincare}
\left\{ Q^\alpha , \bQ_\beta\right\} = - \frac{\im}{2} \, \delta^{\alpha}_{\beta} \mathcal{H}, , \quad \left\{ Q^{\alpha}, Q^{\beta}\right\}=\left\{\bQ_{\alpha}, \bQ_{\beta} \right\}=0.
\ee
To this end, following \cite{WDVVShort}, we   firstly equip  the Riemann manifold $(M_0, g_{ij}(x)dx^idx^j)$ with  the
third-rank symmetric tensor $F_{ijk}(x)dx^idx^jdx^k$
which obeys the equations \eqref{mWDVV0}.

 The first equation  in \eqref{mWDVV0} defines the well-known  Codazzi tensor,
while the second equation  could be viewed as a generalization of
Witten-Dijkgraaf-Verlinde-Verlinde equation \cite{WDVV} 
to Riemann manifolds, and was referred
as {\sl curved WDVV equation} in \cite{WDVVShort,2,kozyrev}.

To construct the supersymmetric mechanics with nontrivial potential we should define on $(M_0, g_{ij}dx^idx^j, F_{ijk}(x)dx^idx^jdx^k)$ the closed one-form obeying the following compatibility condition
\be
W^{(1)}=W_i(x)dx^i\; : \quad dW^{(1)}=0,\qquad W_{i;j} + F_{ijk} g^{km} W_{m} =0.
\label{mW}\ee
Clearly, it can be locally presented as an exact one-form $W^{(1)}=dW(x)$, with the locally  defined  function $W(x)$  called ``prepotential".

With these objects at hands we can construct the $\mathcal{N}=4$ supersymmetric mechanics defined by the following supercharges and Hamiltonian \cite{2}
\bea\label{Q}
& Q^\alpha = p_i \,\psi^{i\alpha } + \im W_i\, \psi^{ i\alpha} +\im F_{ijk}\, \psi^{i\beta}\, \psi^j_{\beta}\, \bpsi^{k\alpha},
\qquad  \bQ _\alpha = p_i\, \bpsi^i_\alpha  - \im W_i\, \bpsi^i_\alpha +\im F_{ijk}\, \bpsi^i_\beta\, \bpsi^{j\beta}\, \psi^k_\alpha\; ,
&\\
&\mathcal{H}= g^{ij} { p}_i { p}_j + g^{ij}  W_i  W_j +4   W_{i;j} {\psi}^{i\alpha} {\bpsi}_{\alpha}{}^j -4\big[  F_{kmj;i} + R_{imjk} \big] {\psi}^{i\alpha}  {\bpsi}_{\alpha}{}^m\,{\psi}^{j\beta} {\bpsi}_{\beta}{}^k &\label{H}
\eea
 It follows from  \eqref{mWDVV0}  that there exists the special coordinate frame where the metrics (and, respectively, Christoffel symbols and Riemann  tensor) takes the form
 \cite{kozyrev}
\be\label{RK1}
g_{ij} = \frac{\partial^2{\cal A}}{\partial x^i \partial x^j} \; : \qquad \Gamma_{ijk} = \frac{1}{2}\frac{\partial^3 A(x) }{\partial x^i \partial x^j \partial x^k}, \quad  R_{ijkm} = \Gamma_{imp}g^{pq}\Gamma_{qjk} - \Gamma_{ikp}g^{pq}\Gamma_{qjm}.
\ee
{}From the last equation it becomes clear that the  choice $F_{ijk}=\pm \Gamma_{ijk}$
solves curved  WDVV equations \eqref{mWDVV0}. Then, solving the equations   \eqref{mW} we get the two sets of solutions
\be
 \left( F_{ijk} = - \Gamma_{ijk},\;  W = \sum_i c_i \frac{\partial {\cal A}}{\partial x^i}\right), \qquad
\left (F_{ijk} =  \Gamma_{ijk},  \; W = \sum_i c_i x^i\right), \quad {\rm with }\quad c_i = const. \label{RKPot3}
\ee
The first solution is the one   obtained in \cite{AP1}. The second solution could be transformed to the first one by the  Legendre  transformation
\be
x^i\to u_i =\partial_i\mathcal{A}( x),  \qquad \mathcal{A}(x)\to \tilde{\mathcal{A} } (u)= (u_i x^i-\mathcal{A}(x))\vert_{u_i=\partial_i\mathcal{A}(x)}.
\label{legendre}\ee
In this coordinate frame the system \eqref{Q},\eqref{H} coincides with  $N$-dimensional $\mathcal{N}=4$ supersymmetric mechanics
 constructed  by using the $N$ scalar supermultiplets \cite{AP1} (the respective system with single supermultiplet was investigated in  \cite{IKL}).

However, in many cases it is more convenient to solve the  equations \eqref{mWDVV0},\eqref{mW} in other frames.
Below we will exemplify this by presenting their solutions  on $so(N)$-invariant special Riemann manifolds.

\subsection*{$SO(N)$-invariant  Riemann manifolds}
Let us consider  curved  WDVV  and potential equations \eqref{mWDVV0}, \eqref{mW}  on  isotropic  ($so(N)$-invariant)  spaces with the metric represented in conformal  flat form
\be\label{CFlat1}
g_{ij}dx^idx^j =\sum_{i=1}^N g(r) {dx^idx^i},\qquad {\rm where}\quad r^2={\sum_{i=1}^N x^ix^i}.
\ee

Let us show that these manifolds  always admit nontrivial solutions.

Indeed, let  $F_{(0)ijk}(x)$ and $W_{(0)}(x)$ be the solutions of the WDVV  and potential equations  on the Euclidian space which obey some additional condition
\be\label{flatWDVV}
 F_{(0)ikp}F_{(0)pjm}=F_{(0)jkp}F_{(0)pim},\quad \partial_i \partial_j W_{(0)}  + F_{(0)ijk}\partial_k W_{(0)}=0,
\ee
 { with}
 \be
 \sum_{i=1}^Nx^i F_{(0)ijk} =  \delta_{jk},
\quad \sum_{i=1}^Nx^i \partial_i W_{(0)} = \alpha_0,
\ee
where $F_{(0)ijk}=\frac{\partial^3 F_{(0)}}{\partial x^i\partial x^j\partial x^k}$ and $\alpha_0$ is   some constant.
The variety of  pairs $(F_{(0)}, W_{(0)})$ obeying these equations was presented  in \cite{GLP}.

These flat solutions can be lifted to the  solutions of
{\sl curved  WDVV and potential  equations} on   isotropic  spaces  as follows (we use here the notation which slightly differs  from that in \cite{WDVVShort,2})
\be
F^{\sigma}_{\kappa|ijk}=g(r)\left({ F_{(0)ijk}}+\Gamma(r) \frac{\delta_{ij}x^k   +
 \delta_{jk}  x^{i} +\delta_{ik} x^{j}}{r^2} -A(r)\frac{x^{i} x^{j} x^{k}}{r^4}\right),
\label{Fkappa}\ee
where
\be\label{AB}
\Gamma (r)=\frac r2 \frac{d\log g}{dr},\quad
A(r)=2\Gamma-
\frac{r\Gamma'/2}{\Gamma+1}.
\ee
Corresponding  solution   of curved potential equation  and the respective Hamiltonian are given by the expressions
\be\label{solpot1}
W = W_{(0)}  - \alpha_0 \int \frac{\Gamma}{1+\Gamma} \frac{dr}{r} \;,\qquad\Rightarrow\qquad
H=g^{-1}(r)\sum_{i=1}^N\left(p_i p_i+ W_{(0)i} W_{(0)i}\right) -\frac{ 2\alpha^2_0}{rg(r)} \left(1-\frac{1}{(1+\Gamma)^2}  \right) .
\ee
Note that  the  "curved" counterpart of the initial Hamiltonian  yields an additional potential term with coupling constant $\alpha_{0}^2$. In the
 particular case of sphere and two-sheet hyperboloid (pseudosphere), when  $g=(1+\epsilon  r^2)^{-2}$ (with  $\epsilon=1$  corresponding  to the sphere and   $\epsilon=-1$
  to the pseudosphere),   it  coincides   with the
  potential  of the
 superintegrable  (pseudo)spherical generalization  of harmonic oscillator known as
Higgs oscillator \cite{higgs}.

Thus, with the specific choice of  initial prepotential  $W_{(0)}(x)$  we can construct  $\mathcal{N}=4$ supersymmetric superintegrable  deformations  of  Higgs oscillator.
For example, the  choice $
W_{(0)} = \sum_{i=1}^N \alpha_i \log x^i$, $F_{(0)}= \frac{1}{2}\sum_{i=1}^N (x^{i})^2 \log x^i$,
 yields  superintegrable (pseudo)spherical deformation of $N$-dimensional oscillator with extra centrifugal terms  (which is also  known as Rosochatius system)
 with additional restriction to the oscillator frequency \cite{2}
\be\label{ex1c}
H_{Ros}=(1+\epsilon r^2)^2\left (\sum_{i=1}^Np^2_i + \sum_i\frac{\alpha_i^2}{x^{i2}}+\epsilon
\frac{ 4\left(\sum_i \alpha_i\right)^2}{\left( 1-\epsilon r^2 \right)^2}\right).
\ee
Taking the solutions of  \p{flatWDVV} corresponding to the three-particle rational Calogero model \cite{Wyllard}, we will get  the following (pseudo)spherical Hamiltonian

\be\label{ex2c}
H_{3Calogero} =  \left(1+\epsilon r^2\right)^2 \left(\sum_{i=1}^3p^2_i+ \sum_{i>j=1}^3\frac{2 g^2}{(x_i-x_j)^2} +\epsilon  \frac{36 g^2}{\left(1 -\epsilon r^2\right)^2}\right).
\ee
It is a particular  case of superintegrable (pseudo)spherical Calogero-Higgs oscillator \cite{Armen3}.

\setcounter{equation}{0}
\section{ $\mathcal{N}=8$ mechanics on special  K\"ahler manifolds}
In this Section we generalize the system  presented in \cite{BKN}, and construct, on the special K\"ahler manifolds of the rigid type,
 the $(N|4N)$-(complex)dimensional mechanics with potential term,  which  possesses the $\mathcal{N}=8$ supersymmetry
 \be
  \left\{ Q_{i \alpha}, Q_{j\beta} \right\}=\left\{ \bQ_{i \alpha}, \bQ_{j \beta}\right\} =0, \qquad \left\{ Q_{i \alpha}, \bQ_{j \beta}\right\} = -\im\, \epsilon_{i j} \epsilon_{\alpha \beta} \mathcal{H} .\label{QQ}
 \ee
For this purpose we  define the $(2N|4N)_{\mathbb{C}}$-dimensional phase superspace equipped by the
supersymplectic structure
\be
\Omega=d\pi_a\wedge dz^a+ d{\bar\pi}_{\bar a}\wedge d{\bar z}^a
- R_{a\bar bc\bar d}\eta^c_{i\alpha}\bar\eta^{d|i\alpha}
dz^a\wedge d{\bar z}^b+ g_{a\bar b}D\eta^a_{i\alpha} \wedge
D{\bar\eta}^{b| i\alpha}, \qquad D\eta^a_{i\alpha} =d\eta^a_{i\alpha}+\Gamma^a_{bc}\eta^b_{i\alpha}
dz^c,
\label{ss}\ee
with fermionic variables $\eta^a_{i\alpha}$ and $\b{\eta}{}^{\b{a}}_{i \alpha}$  related as follows
$
 \left( \eta^a_{i \alpha}\right)^\dagger = \b{\eta}{}^{\b{a} i \alpha}
$. Here $\alpha, i=1,2$ are $su(2)$ indices
which are  raised and lowered as follows:
 $A_\alpha = \epsilon_{\alpha\beta}A^\beta,  A^\alpha=\epsilon^{\alpha\beta} A_\beta$,  $A_i = \epsilon_{ij}A^j,  A^i=\epsilon^{ij} A_j$,with $\epsilon_{12}=\epsilon^{21}=1$.

This supersymplectic structure is  manifestly invariant under the coordinate transformation
\be\label{ct}
{\widetilde z}^a={\widetilde z}^a(z), \qquad {\widetilde\pi}_a=\frac{\partial z^b}{\partial{\widetilde z}^a}\pi_b, \qquad{\widetilde\eta}^a_{i\alpha }=\frac{\partial{\widetilde z}^a}{\partial z^b}\eta^b_{i\alpha },
\ee
i.e. $\eta^a_{i\alpha}$ transforms as  $dz^a $.

The Poisson brackets corresponding to \eqref{ss} are defined by the relations
\be\label{PBnew}
 \left\{ \pi_a , z^b\right\} =\delta_a^b, \qquad \left\{\pi_a, \eta^b_{i\alpha}\right\}= - \Gamma^b_{ac}\, \eta^c_{i\alpha}, \qquad
 \left\{\pi_a, \b{\pi}_{\bar b}\right\}= \im\, R_{a\b{b}c\b{d}}\, \eta^c_{i\alpha} \b{\eta}^{\b{d} i \alpha},\quad
\left\{\eta^a_{i \alpha}, \b{\eta}^{b j \beta}\right\}= - \im g^{a\b{b}}\,\delta_i^j \,\delta_\alpha^\beta\;.
\ee

{}For the construction of supersymmetric mechanics with nonzero potential we have to equip the K\"ahler  manifold
by the closed holomorphic one-form
\be
U^{(1)}=U_a(z)dz^a\;, \quad U_a=\frac{\partial U(z)}{\partial z^a},
\label{u1}\ee
where  $U(z)$ is a  locally defined holomorphic function which is called ``prepotential".

With these ingredients at hand we can construct
the  $\mathcal{N}=8$ supersymmetric mechanics {\bf \sl with} potential term.
Having in mind the structure of supercharges of the $\mathcal{N}=4$ supersymmetric mechanics on generic K\"ahler manifold \cite{PRDrapid}, and
of the the $\mathcal{N}=8$ supersymmetric mechanics ({\sl without} potential term) on special K\"ahler manifolds \cite{BKN},
 we choose the following anzats for supercharges:
\be\label{Qnew}
Q_{i\alpha}  =  \pi_a \eta^a_{i\alpha} +  \bU_{\b{a}}\,T_{\alpha}{}^\beta \b{\eta}{}^{\b{a}}_{i \beta} +
\frac{\im}{3} \b{f}_{\b{a}\b{b}\b{c}}\,\b{\eta}{}^{\b{a}}_{i \beta}\b{\eta}{}^{\b{b} j \beta}\b{\eta}{}^{\b{c}}_{j \alpha}, \qquad
\bQ_{i\alpha} = \b{\pi}_{\b{a}}\, \b{\eta}{}^{\b{a}}_{i \alpha} - U_a \, T_\alpha{}^\gamma \eta^a_{i \gamma}+
\frac{\im}{3} f_{abc}\, \eta^a_{i\beta} \eta^{b j \beta} \eta^c_{j \alpha},
\ee
where the matrix $T_{\alpha}^{\beta}$ collects the parameters which control the explicit breaking of the $su(2)$ symmetry realized on the Greek indices, and, without loss of generality, is parameterized by the {two  angle-like parameters $\alpha_0,\beta_0$}
\be\label{T}
T_{\alpha}^\beta = \left( \begin{array}{cc} \cos\alpha_0 &  {\rm e}^{\im\beta_0}\sin\alpha_0 \\
 {\rm e}^{-\im\beta_0}\sin\alpha_0 & - \cos\alpha_0\end{array}\right).
\ee
One should stress that it is impossible to introduce the interaction which preserved both $su(2)$ symmetries
(realized on the Greek and Latin indices from the middle of alphabet ($i,j,k$)). However the simultaneous breaking of both these symmetries results in the appearance of the central charges in the super Poincar\'{e}
algebra \cite{BKNS}.

The components of (anti)holomorphic symmetric tensors $f_{abc}(z)$ and $\b{f}{}_{\b{a}\b{b}\b{c}}(\bar z)$ have to obey
the constraints  \eqref{sKrt}, and $U_a$ and $\bU_{\b{a}}$ were defined in \eqref{u1}.

Then, taking their Poisson brackets we find that these supercharges span the $\mathcal{N}=8$ Poincar\'{e} superalgebra \eqref{QQ} if  $U_a$ and $\bU_{\b{a}}$ obey the equations
\be
U_{a;b}-f_{abc}g^{\b{d}c}\bU_{\b{d}}=0.
\label{potc}
\ee
with  $U_{a;b} = U_{a,b}-\Gamma_{ab}^c U_c$.
In such a case, the Hamiltonian reads
 \bea\label{Hnew}
 &\mathcal{H}&= g^{a\b{b}} (\pi_a\b{\pi}_{\b{b}} + U_a\bU_{\b{b}}) -
 \frac{\im}{2} U_{a} g^{a\b{e}}\b{f}_{\b{e}\b{b}\b{c}}\b{\eta}_{\;i}^{\b{b}\alpha}T_\alpha{}^\beta \b{\eta}_{\;\beta}^{\b{c}i} -
  \frac{\im}{2} \b{U}_{\b{a}} g^{\b{a}e }f_{e b c}{\eta}_{\;i}^{{b}\alpha}T_\alpha{}^\beta {\eta}_{\;\beta}^{{c}i} -\nn\\
 &&
 \frac{1}{12} f_{abc;d} \eta^{a\,i\rho} \eta^b_{\;i\gamma}\eta^{c\, j\gamma}\eta^d_{j \rho} -
 \frac{1}{12} \b{f}_{\b{a}\b{b}\b{c};\b{d}} \b{\eta}^{\b{a}\, i \rho} \b{\eta}^{\b{b}}_{\;i\gamma}\b{\eta}^{\b{c}\, j\gamma}\b{\eta}^{\b{d}}_{\;j \rho} -
 \frac{1}{4} f_{abe} g^{\b{e}' e}\b{f}_{\b{e}'\b{c}\b{d}}\, \left(
  \eta^{a i }_{\;\alpha}\eta^{b}_{\;i \beta} \b{\eta}^{\b{c}j \alpha\,} \b{\eta}^{\b{d}\beta\,}_{\;j}+
  \eta^{a\alpha }_{\;i} \eta^{b}_{\; j \alpha} \b{\eta}^{\b{c}j \beta\,} \b{\eta}^{\b{d}i\,}_{\;\beta}\right).
 \label{Ham1}\eea

The equations \eqref{sKrt} can  be expressed in the distinguished coordinate frame via single holomorphic function (``Seiberg-Witten potential") $\mathcal{F}(z)$
 (see, e.g. \cite{fre})
\be
 g_{a\bar b}= \mathbb{R} {\rm e}\;\partial_a\partial_b \mathcal{F}(z),\qquad
\Gamma_{\bar abc}=\partial_a\partial_b\partial_c\mathcal{F}(z),\quad \Rightarrow \quad f_{abc}=\Gamma_{\bar a bc}
\label{sccf}.\ee

 In this coordinate frame the  equation \eqref{potc} looks as follows
\be
{\partial_a\partial_b}U- \left(\partial_d U + \partial_{\bar d}\bU      \right)g^{\bar d c}\partial_a\partial_b\partial_c\mathcal{F}=0
\label{potcl}\ee
{}From this equation we immediately get the following solution
\be
U(z)=\sum_{a=1}^N \left(m^a\partial_a \mathcal{F}(z) +\im n_a z^a\right),
\label{csol}\ee
with $m^a, n_a$ being real constants.\\

The bosonic part of constructed  $\mathcal{N}=8$ supersymmetric mechanics respects ``T-duality" transformation, which is the complex counterpart of Legendre transformation \eqref{legendre}
\be
z^a\to  u_a=\partial_a\mathcal{F},  \quad \mathcal{F}(z)\to \mathcal{\tilde F} (u)=      \left( z^au_a -\mathcal{F}(z)\right)\vert_{u_a=\partial_a\mathcal{F}}.
\ee
For the potential it reads
\be
U(z)=\sum_{a=1}^Nm^a\partial_a\mathcal{F}(z)+\im n_az^a,\quad \to\quad {U}(u)=\sum_{a=1}^N m^au_a +\im n_a \partial^a\mathcal{{\tilde F}}(u)
\ee
The extension of duality transformation to the
whole phase superspace is as follows \be
(z^a ,\pi_a,\, \eta^{a}_{i\alpha})\to ( u_a, p^a, \xi^{a}_{\im\alpha}),\qquad{\rm where}\quad
 {u}_a=\partial_a \mathcal{F}(z),\quad
p^a\frac{\partial^2 \mathcal{F}}{\partial z^a\partial z^b}=-\pi_b
 ,\quad \theta_{a\;i\alpha}=
\frac{\partial^2 \mathcal{F}}{\partial z^a\partial z^b}\xi^{b}_{\;i\alpha}.
\ee
Looking  back at the presented model of $\mathcal{N}=8$ supersymmetric mechanics we can observe   many similarities
 with $\mathcal{N}=4$ supersymmetric mechanics described in the previous Section,  which prompts us to consider it as a  complex counterpart of  the latter one.
In particular,  the notion of "special K\"ahler manifold of the rigid type" \eqref{sKrt} can be viewed as the complex analog of
 curved WDVV equations \eqref{mWDVV0}, and  as well as the restriction on the prepotential $U(z)$  can be viewed as  a complex counterpart of those to the real one \eqref{WF}.
In both cases there exist special coordinate frames where the metrics and the respective third-rank tensors are expressed via single  function, cf. \eqref{RK1} and  \eqref{sccf}.
Further visible similarities can be noticed comparing \eqref{csol}  and \eqref{RKPot3}.

However, the requirement of  ``special K\"ahleriality"  \eqref{sKrt} is more restrictive than  \eqref{mWDVV0}. Say, special K\"ahler manifold of the rigid type necessarily has a negative curvature,
 while ``curved WDVV equations" does not yield such restriction; ``curved WDVV equations"
admit the nontrivial  solutions on the generic $so(N)$-invariant Riemann manifolds (including   $N$-dimensional spheres and hyperboloids).
In contrast to this,  complex projective spaces (and their non-compact counterparts) cannot be equipped by the structure of special K\"ahler manifold.
Moreover, it seems that special K\"ahler metrics  could possess the $U(N)$ isometry only in the simplest case $N=1$ to be considered in the next Section.

\section{Two-dimensional systems}
  In this Section we construct the one-(complex)dimensional special K\"ahler manifolds  which is invariant under $U(1)$-transformation $z\to{\rm e}^{\im\lambda}z$,  and then find the potentials admitting $\mathcal{N}=8$ supersymmetric extension.

Choosing the metric $g$ to be the function of $z \bar z$ only, i.e. putting $g=g(z\bar z)dzd{\bar z}$ one may
explicitly solve the second equation in \eqref{sKrt} as
\be
g(z\bar z)dzd\bar z= \left(c_1(z\bar z)^{n_1}+c_2(z\bar z)^{n_2}\right)dzd\bar z,\qquad f(z)[dz]^3=\sqrt{-c_1c_2 }(n_1-n_2) z^{n_1+n_2-1}[dz]^3, \quad c_{1} c_{2} <0.
\label{cm}\ee
Corresponding K\"ahler potential reads
\be
K(z,\bar z)=\frac{c_1(z\bar z)^{n_1+1}
   }{\left(n_1+1\right){}^2}+\frac{c_2(z\bar z)^{n_2+1}
   }{\left(n_2+1\right)^2}
\ee
Then, performing transformation  $ \frac{\sqrt{c_1}}{n_1+1}z^{n_1+1}\to z$, we can simplify  these structures as follow
\be
ds^2=(1-\kappa^2(z\bar z)^{m})dz d\bar z, \quad f(z)[dz]^3={\kappa}m z^{m-1}\,[dz]^3,\quad {\rm with}\quad |z|\in[0,{\kappa^{-1/m}}).
\label{sKmf}\ee
The Christoffel symbol and the  Riemann curvature  are
\be
\Gamma^{1}_{11}=-\frac{\kappa^2\, m\, z^{m-1}\bar z^{m}}{1-\kappa^2(z\bar z)^m},
\qquad
R_{1\bar 11\bar 1}=-\frac{\kappa^2\, m^2\,(z\bar z)^{m-1}}{1-\kappa^2(z\bar z)^m}.
\ee
For this special case
the potential equation \eqref{potc} takes the form
\be
U''+\frac{\kappa^2\, m\, z^{m-1}\bar z^{m}}{1-\kappa^2\, (z\bar z)^m} U'-\frac{{\kappa}\, m\,  z^{m-1}}{1-\kappa^2\, (z\bar z)^m}\bar U'=0.
\ee
Then we obtain
\be
 \frac{d{\bar U}(\bar z)}{dz}=\frac{d}{d z}\Big(\frac{1-\kappa^2\, (z\bar z)^m}{{\kappa}\, m\, z^{m-1}} U''(z)+{\kappa}\, {\bar z}^m\, U'(z) \Big)=0,
\ee
From this equation we immediately get the  solution
\be
U'(z)= \kappa a z^{m}+{\bar a}
\label{U1}\ee
with $a$ being an  arbitrary complex constant.

Thus, the  one-(complex)dimensional   $\mathcal{N}=8$ supersymmetric mechanics is defined  by the following bosonic
Hamiltonian
\be
H_{\kappa,m,a}=\frac{\pi\bar\pi +|\kappa a z^m+{\bar a}|^2}
{1-\kappa^2 (z\bar z)^m},\qquad{\rm with } \quad \{\pi, z\}_0=\{\bar\pi,\bar z\}_0=1,\quad  \{\pi,\bar\pi\}_0=\{z,\bar z\}_0=0
.
\label{bosH}\ee
The presence of nonzero potential breaks the  kinematical $U(1)$- symmetry $z\to{\rm e}^{\im\lambda}z, \pi\to{\rm e}^{\im\lambda}\pi $.
But in the
 free particle case $a=0$ the hamiltonian  becomes manifestly invariant  under this   transformation
 and thus defines the   integrable system
\be
H_{\kappa,m,0}=\frac{\pi\bar\pi }{1-\kappa^2 (z\bar z)^m}\, \quad J=\im \left( z\pi- \bar z \bar\pi\right)\;:\qquad \{H_0,J\}_0=0
\label{bosH0}\ee
where $J$ is the generator of $U(1)$-symmetry.

 However, for the specific values of $m$ the system could have the hidden symmetries.
The simplest example corresponds to the $m=-2$ case.
\begin{itemize}
 \item $m=-2$

In this case the  Hamiltonian \eqref{bosH} admits the separation of variables in the polar coordinates
\be
z=r{\rm e}^{\im\varphi},\;  \pi=\frac{{\rm e}^{-\im\varphi}}{2}\left(p_r-\frac{\im p_{\varphi}}{r}\right)\; :\;
H_{\kappa,-2,a}=\frac{p^2_r+|a|^2(1+\frac{\kappa^2}{r^4}) }{4(1-\frac{\kappa^2}{r^2})}+\kappa \frac{ p^2_\varphi+\kappa|a|^2\cos(\varphi+{\rm arg}\; a) }{4(r^2-{\kappa^2})}.
\ee
 which allows immediately find the  quadratic constant of motion
\be
H_{\kappa,-2,a}=\frac{\pi\bar\pi +|\kappa a z^{-2}+ \bar a|^2}{1-\frac{\kappa^2}{ |z|^2}},\quad
I=p^2_\varphi+2\kappa |a|^2\cos(\varphi+{\rm arg}\; a)  =\left( z\pi- \bar z \bar\pi\right)^2 - 4\kappa \frac{{\bar a}^2 z^2 +a^2 {\bar z}^2}{z {\bar z}}.
\label{uni}\ee
\end{itemize}
To find additional values of the parameter $m$ leading to the  (super)integrable systems, one has to do the following.
Fixing the energy surface of the Hamiltonian \eqref{bosH} one may re-write it  as
\be
\pi\bar\pi +\kappa^2(|a|^2+E_{\kappa,m,a})|z|^{2m}+ \kappa a^2 z^m+{\kappa\bar a}^2{\bar z}^m =E_{\kappa,m,a}-|a|^2
\ee
From this expression we immediately deduce that for $m=1$ it coincides with the energy surface of the two-dimensional oscillator interacting with linear electric field which could be absorbed by the trivial sift of complex coordinate $z$,  while for the $m=-1/2$ it can be easily transform to the $m=1$ case by the Bohlin-Levi-Civita  transformation $z={\widetilde z}^2$ which relates the energy surfaces of two-dimensional oscillator and Coulomb problem. Hence,  for  the particular values of $m=1,-1/2$ the Hamiltonian \eqref{bosH} possesses two functionally independent  constants of motion and hence becomes superintegrable.
Let us consider these cases in the full details.

 \begin{itemize}
\item $m=1$

In this particular case the Hamiltonian \eqref{bosH}
takes a form
\be
H_{\kappa,1,a}=\frac{\pi\bar\pi +|\kappa a z+\bar a|^2}
{1-\kappa^2 |z|^2}.
\label{h1}\ee
 It possesses  the hidden symmetry given by the
 deformed $U(1)$-generator $J$ presented in \eqref{bosH0}
 \be
 J_{\kappa,1}= \im\left[ \left(z+\frac{\bar a^2}{\kappa \left(|a|^2+ H_{\kappa,1,a}\right)}\right)\pi-\left({\bar z}+\frac{a^2}{\kappa \left( |a|^2+H_{\kappa,1,a}\right)}\right) \bar\pi\right]\; \\
 \ee
  and by the
  complex constant of motion
\be
F_\kappa = \pi^2 +\kappa^2 \left( |a|^2 + H_{\kappa,1,a}\right) \left({\bar z}+\frac{a^2}{\kappa \left( |a|^2 +H_{\kappa,1,a}\right)}\right)^2,
\label{osc}\ee
which can be interpreted as a deformation of the so-called Fradkin tensor written in complex coordinates  $z=(x_1+\im x_2)/\sqrt{2}$ and conjugated momentum.

They form the nonlinear algebra
\be
\left\{J_{\kappa, 1}, F_\kappa\right\} = 2 \im F, \quad \left\{J_{\kappa,1}, {\bar F}_\kappa\right\} = - 2 \im {\bar F}_\kappa, \quad
\left\{F_\kappa, {\bar F}_\kappa\right\} = 4 \im \kappa^2 \left( |a|^2 + H_{\kappa,1,a }\right) J_{\kappa,1}.
\label{aosc}\ee

For emphasizing    the relation of this system with oscillator, let us re-write the Hamiltonian \eqref{h1} it as follows
\be
H_{\kappa,1,a }= H^{\kappa}_{osc}+\frac{|\omega|^2}{2\kappa^2}, \qquad H^{\kappa}_{osc}= \frac{\pi\bar\pi +|\omega|^2 z\bar z + {\bar E}z+ {E}\bar z}{1-\kappa^2 (z\bar z)},\quad
\omega:=\sqrt{2}\kappa |a|,\quad E:= {\kappa {\bar a}^2}.
\ee
The function in numerator can be interpreted as a two-dimensional isotropic oscillator with the frequency $|\omega|$ interacting with electric field ${\bf E}=(E_1,E_2)$ with $E=( E_1+\im E_2)/2$.
The parameters $\kappa, {a} $ can be expressed  via $\omega, E$ as follows
\be
\kappa=\frac12 \frac{|\omega|^2}{|E|},\quad a=\sqrt{2}{\rm e}^{-\im\frac{{\rm arg} E}{2}}\frac{|E|}{|\omega|}.
\ee

\item $m=-1/2$

In this  case the Hamiltonian  \eqref{bosH} acquires the  form
\be
H_{\kappa,-1/2,a}=\frac{\pi\bar\pi +|\kappa a \frac{1}{\sqrt{z}}+\bar a|^2}
{1-\frac{\kappa^2}{|z|}}.
\label{h-}\ee
It possesses   the hidden symmetry given by the
 deformed $U(1)$-generator $J_{\kappa,-1/2}$
\be
 J_{\kappa,-1/2} = 2 \im \left[\left( z- \frac{\kappa a^2\sqrt{z}}{H_{\kappa,-1/2,a}-|a|^2}\right) \pi-
\left( {\bar z}- \frac{\kappa{ \bar a}^2\sqrt{\bar z}}{H_{\kappa,-1/2, a}-|a|^2}\right) {\bar \pi} \right],
\ee
and by the complex constant of motion being the deformation of two-dimensional Runge-Lenz vector  ${\bf A}=(A_1,A_2)$ with $A_\kappa=( A_1+\im A_2)/2$:
\be
A_\kappa = z \pi^2 -\left(H_{\kappa,-1/2, a} - |a|^2\right)\left( \sqrt{\bar z} -\frac{\kappa {\bar a}^2}{H_{\kappa,-1/2, a}-|a|^2}\right)^2.
\ee
They form the non linear algebra
\be
\left\{J_{\kappa, -1/2}, A_{\kappa}\right\} = 2 \im A_{\kappa}, \quad \left\{J_{\kappa,-1/2}, {\bar A}_\kappa\right\} = - 2 \im {\bar A}_\kappa, \quad
\left\{A_\kappa, {\bar A}_\kappa\right\} = - \im \left( H_{\kappa,-1/2,a} - |a|^2\right) J_{\kappa,-1/2}.
\ee
The Hamiltonian \eqref{h-} can be interpreted as a  deformation of the  two-dimensional Coulomb problem
perturbed by the potential $\delta V=k \left( \frac{a^2}{\sqrt{z}}+\frac{{\bar a}^2}{\sqrt{\bar z}}\right)$.
\end{itemize}

The bosonic Hamiltonian $H$ \eqref{bosH} possesses the following duality transformation:
\be
H_{\kappa,m,a}=\frac{\pi\bar\pi +|\kappa a z^m+\bar a|^2}
{1-\kappa^2 |z|^m} = - \frac{\tilde{\pi}\bar{\tilde\pi} +|\tilde\kappa {\tilde a}] {\tilde z}^{\tilde m}+
\bar\tilde{a}|^2}
{1-{\tilde\kappa}^2 |\tilde z|^{\tilde m}} = - H_{\tilde\kappa, \tilde m,\tilde a},
\ee
where the variables are related as
\be
z=\frac{{\widetilde\kappa}{\widetilde z}^{{\widetilde m}+1}}{{\widetilde m}+1},\qquad \pi=\frac{\widetilde\pi}{{\widetilde\kappa}{\widetilde z}^{\widetilde m}},
\label{bohlin}\ee
and the following constraints on the parameters are imposed
\be
 (m+1)({\widetilde m}+1)=1,\quad \kappa{\widetilde\kappa}^{m+1}=|{\widetilde m}+1|^m, \quad  {\tilde a}=\bar a .
\label{equivalence}\ee
To be self-consistence, the transformations \eqref{bohlin} should be supplied by the changing of the admitted
values of the coordinates
\be
\mbox{ From    }  |z|\in [0,\kappa^{-1/m})\quad \mbox{to    } \quad |\widetilde z|\in [{\widetilde\kappa}^{-1/{\widetilde m}}, \infty ).
\ee
\\
Explicitly, the   supercharges  of the $\mathcal{N}=8$ supersymmetric extensions of the presented bosonic systems read
\be
Q_{i\alpha }=\pi \eta_{i\alpha}+(\kappa\bar a\bar{z}^{m}+a)T_\alpha^\beta\bar\eta_{i \beta}+\frac{\im }{3}{\kappa}m{\bar z}^{m-1} \bar\eta_{i\beta}\bar\eta^{j \beta}\bar\eta_{j \alpha},
\label{Q1}\nn\ee
while   the  Hamiltonian has the form
\bea
\mathcal{H}_{\kappa,m,a} &=& \frac{\pi \bar\pi+ |\kappa a z^m+\bar a|^2}{ 1 - \kappa^2(z\bar z)^m}+
 \im \frac{ m{\kappa} z^{m-1} (a+ \kappa\bar a {\bar z}^m)}{2(1- \kappa^2(z\bar z)^m)} \eta_i^\alpha T_\alpha^\beta \eta^i_{\beta}-
\im  \frac{ m{\kappa} {\bar z}^{m-1} (\bar a+ \kappa a z^m)}{2(1- \kappa^2(z\bar z)^m)}  \bar\eta_i^\alpha T_\alpha^\beta \bar\eta^i_{\beta}- \nn\\
&& \kappa m\frac{m-1 +\kappa^2 (1+ 2 m)(z\bar z)^m}{12 \left(1 - \kappa^2(z\bar z)^m\right)}\left( z^{m-2} \eta_{i \alpha} \eta^{i\beta} \eta_{j \beta} \eta^{j \alpha}+
{\bar z}^{m-2} \bar\eta_{i \alpha} \bar\eta^{i\beta} \bar\eta_{j \beta} \bar\eta^{j \alpha}\right)- \nn \\
&&\frac{\kappa^2 m^2(z\bar z)^{m-1}}{4 \left(1-\kappa^2(z\bar z)^{m} \right)} \left( \eta^i_\alpha\eta_{i\beta}\bar\eta^{j\alpha}\bar\eta_j^\beta +\eta_i^\alpha \eta_{j\alpha}\bar\eta^{j\beta}\bar\eta^i_\beta\right).\label{shi}
\eea
The  $U(1)$-transformation $z\to {\rm e}^{\im\lambda}z$ extended  to the supersymplectic structure \eqref{ss} looks as follows
\be
z\to{\rm e}^{\im\lambda}z,\pi\to{\rm e}^{-\im\lambda}\pi, \eta_{i\alpha} \to{\rm e}^{\im\lambda}\eta_{i\alpha}.
\label{finite}\ee
It is defined by the generator
\be
\mathcal{J}=\im \left( z\pi- \bar z \bar\pi \right) -  \frac{\partial^2 h(z\bar z)}{\partial z\partial\bar z}\eta^{i\alpha}\bar\eta_{i\alpha},\qquad h(z,\bar z)=z\bar z - \frac{\kappa^2(z\bar z)^{m+1}}{m+1}
\label{Js}\ee
where $h(z,\bar z)$ is the Killing potential  for  $U(1)$-isometry.

It is seen that the supercharges \eqref{Q1} and the Hamiltonian \eqref{shi}  are not invariant  under this transformation for the generic $m$  even for  $a=0$.
It is not surprising, since the  third-order tensor $f(a)[dz]^3$ in \eqref{sKmf} is invariant  under $U(1)$-transformation $z\to {\rm e}^{\im\nu}$ only for the $m=-2$, while  the one-form  \eqref{U1} is not  $U(1)$-invariant  at all. Since $\eta_{i\alpha}$ transforms as $dz $, we conclude that the supercharges and supersymmetric  Hamiltonian fail to be  $U(1)$-invariant in the generic case.
Hence, only in the case  $a=0, m=-2$,  corresponding to the bosonic Hamiltonian \eqref{uni} we can construct the $\mathcal{N}=8$ supersymmetric extension with the  supercharges and Hamiltonian   which are  invariant under transformation \eqref{finite}.

 On the other hand,  the Hamiltonian $\mathcal{H}_{\kappa,m,0}$, in contrast to  supercharges $Q_{(\kappa,m,0)\;i\alpha } $, is invariant under transformation
 \be
 z\to{\rm e}^{\im\lambda}z,\quad \pi\to{\rm e}^{-\im\lambda}\pi, \quad \eta_{i\alpha} \to {\rm e}^{\im\frac{2-m}{4}\lambda}\eta_{i\alpha}.
\label{finm} \ee
Hence, it commutes with the generator
\be
\widetilde{\mathcal{J}}= \im \left( z\pi- \bar z \bar\pi\right) -  \frac{\partial^2 h(z\bar z)}{\partial z\partial\bar z}\eta^{i\alpha}\bar\eta_{i\alpha} + \frac{m+2}{4}g(z\bar z)\eta_{i\alpha}\bar\eta^{i\alpha}\; :\quad\{\widetilde{\mathcal{J}}, \mathcal{H}_{m,0}\}=0,
\ee
where $h(z\bar z)$ is Killing potential \eqref{Js} and  $g(z\bar z)= 1-\kappa^2 (z\bar z)^m$ { is the component of  special Ka\"hler metrics \eqref{cm}}.
%

\section{Concluding remarks}
In this paper  we have constructed the $\mathcal{N}=8$ supersymmetric mechanics with potential term, whose configuration space is a special K\"ahler manifold of the rigid type.
We observed that it can be viewed as a complex counterpart of the recently suggested $\mathcal{N}=4$ supersymmetric mechanics \cite{WDVVShort,2}.
Then we  constructed the $U(1)$-invariant one-dimensional special K\"ahler manifold  and corresponding $\mathcal{N}=8$ supersymmetric mechanics, including $\mathcal{N}=8$ supersymmetric  extensions of superintegrable perturbations of  deformed  two-dimensional oscillator and Coulomb systems considered  in \cite{spaincoul} as particular cases.
It  is an open question whether a $\mathcal{N}=8$  supersymmetric counterparts of the hidden symmetries of theese superintegrable systems  exist.

\acknowledgments

The authors acknowledge
a partial support from the RFBR grant, project No.  18-52-05002 Arm-a (S.K.) and the grants of  the Armenian Committee of Science (A.N., H.S), projects 18RF-002 and 18T-1C106, as well as the   ICTP Network project NT-04.
 The work of H.S. was fulfilled within the ICTP Affiliated Center Program AF-04 and   the Regional Doctoral Program on Theoretical and Experimental Particle Physics Program
 sponsored by VolkswagenStiftung.

\end{document}